\documentclass[
  aps,
  prd,
  onecolumn,
  showpacs,
  superscriptaddress
  ]{revtex4-2}

  \usepackage[dvips]{color}
  \usepackage{natbib}
  \usepackage{latexsym}
  \usepackage{epsfig}
  \usepackage{amsmath}
  \usepackage{amssymb}
  \usepackage{multirow}
  \usepackage{tabularx}
  \usepackage{multirow}
  \usepackage{longtable}
  \usepackage{wasysym}
  \usepackage{hyperref}
  \usepackage{eufrak}
  \usepackage[english, russian]{babel}

  \usepackage{etoolbox}
  \makeatletter
  \patchcmd{\Ginclude@eps}{"#1"}{#1}{}{}
  \makeatother

\begin{document}


  \newcolumntype{C}{>{\centering}X}
  \newcolumntype{Y}{>{\centering\arraybackslash}X}
  \newcolumntype{K}{>{\centering\arraybackslash\arraybackslash}X}

  \newcommand{\A}{\mathfrak{A}}
  \newcommand{\B}{\mathfrak{B}}

  \newcommand{\etal}{\emph{et\,al.}}
  \newcommand{\MC}[3]{\multicolumn{#1}{#2}{#3}}

  \setcounter{MaxMatrixCols}{13}

\title{Projection operator onto spin--$S$ eigenspaces \\ of total and orbital angular momenta}

  \author{M.~I.~Krivoruchenko}
  \email{E-mail address: mikhail.krivoruchenko@itep.ru}
  \affiliation{National Research Centre ``Kurchatov Institute'', 123182 Moscow, Russia}
               
  \date{\today}
  
\begin{abstract}
The Frobenius covariant is used to construct a projection operator onto the spin--$S$ 
eigenspaces associated with the squares of the total and orbital angular momenta. 
The covariant admits two equivalent representations: as a polynomial in powers of the scalar product of 
the spin and orbital angular momentum operators, and as a finite expansion in terms of their respective 
polarization operators. 
A correspondence is established with Villars’ angular momentum projection, used in nuclear structure studies.
\end{abstract}

\maketitle

Projection operators have numerous applications in quantum mechanics, nuclear structure theory, and quantum field theory.
Variational methods are known to typically violate the symmetries of the original Hamiltonian.
In nuclear shell models, the restoration of the translational, Galilean, and rotational symmetries, 
as well as the construction under pairing schemes of states with a well‑defined particle number, 
is accomplished using projection operator techniques \cite{Schmid:2004,Raduta:2015}.
Plane-wave solutions of relativistic wave equations for spin-$1/2$ particles \cite{Berestetskii:1982} and higher-spin
particles (see, e.g., \cite{Cotogno:2020}), including their symmetry properties, 
are described with the help of projection operators. 
This approach enables the derivation of covariant algebraic expressions for transition matrix elements, 
which provide enhanced computational efficiency in modeling scattering processes
\cite{Krivoruchenko:1994,Prokhorov:1999,Ahmadiniaz:2023}.

For systems with spherical symmetry, projection operator techniques offer 
significant, but not fully exploited, potential for describing particle dynamics. 
To illustrate the range of possible applications, we provide three examples.

A first illustrative example is the filling of a vacancy in an atomic electron subshell: $%
(nJLM)\rightarrow (nJLM)^{\prime }+\gamma$, where $n$ is the principal
quantum number, $J$ is the total angular momentum, $M = -J,\ldots, J$ is the magnetic quantum number,
and $L$ is the orbital angular momentum. The electron from the $(nJL)$
subshell migrates to an unoccupied state $(nJLM)^{\prime }$ with the emission of a
photon $\gamma $.

All atomic $\beta$-processes provide additional illustrations, including single and double $\beta$-decays 
as well as electron capture \cite{Simkovic:2021}. To calculate the rates for electron capture 
and bound‑state $\beta$-decay, one sums over the magnetic quantum numbers of the captured electron.

Particle physics also deals with decays of charged particles trapped by nuclei. 
As another illustration, we mention the decays of a spin‑$3/2$ $\Omega ^{-}$-hyperon 
trapped in an atomic orbit \cite{Krivoruchenko:2008}. The calculation of 
$\Omega ^{-}$-decay rates involves averaging over the magnetic quantum numbers.
  
In the non-relativistic approximation, matrix elements of processes with
spherical symmetry take the form 
\begin{equation}
\mathfrak{M}_{fi}\propto \mathrm{tr}[\hat{V}Y_{JM}^{LS}(\mathbf{n}%
)Y_{J^{\prime }M^{\prime }}^{L^{\prime }S^{\prime }\dag }(\mathbf{n}^{\prime
})],  \label{Mfi}
\end{equation}%
where $Y_{JM}^{LS}(\mathbf{n})$ and $Y_{J^{\prime }M^{\prime }}^{L^{\prime}S^{\prime }}(\mathbf{n}^{\prime })$
represent the spherical tensors of the initial- and final-states particles.
Here, the spinor indices are suppressed; $\mathbf{n}$ and $\mathbf{n}^{\prime }$
are the unit vectors in the directions of either the particles' momenta or their positions. 
The operator $\hat{V}$ describes the interaction. 
In the first example
considered, $Y_{J^{\prime }M^{\prime }}^{L^{\prime }S^{\prime }}(\mathbf{n}%
^{\prime })$ appears in the wave function describing the electron shell vacancy. 
Equation (\ref{Mfi}) also applies to electron capture and bound‑state
$\Omega ^{-}$ decay, where the spherical tensors describe the polarization and angular 
distribution of the decay products: the neutrino and either the $\Lambda $-hyperon or the $\Xi ^{0}$-hyperon, respectively.

In practically important cases, when the decay probability is summed over $M$, the spherical tensors can be replaced by an operator
\begin{equation}
\mathcal{\varrho }_{J}^{LS}(\mathbf{n},\mathbf{n}_{\ast})
=\sum_{M}Y_{JM}^{LS}(\mathbf{n})Y_{JM}^{LS\dag }(\mathbf{n}_{\ast }),  \label{rhoLSJ}
\end{equation}%
that projects wave functions of spin-$S$ particles 
onto the simultaneous eigenspaces of $\mathcal{J}^2 = \mathcal{J}^{\mu}\mathcal{J}_{\mu}$ and $\mathcal{L}^2 = \mathcal{L}^{\mu}\mathcal{L}_{\mu}$. 
In what follows, $\mathcal{S}^{\mu}$, $\mathcal{J}^{\mu}$, and $\mathcal{L}^{\mu}$ are the spin, total and orbital angular momentum operators, respectively,
and $\mathcal{J}^{\mu }=\mathcal{S}^{\mu }+\mathcal{L}^{\mu }$.
The expression for the decay probability can be written in the form 
\begin{equation}
\sum_{MM^{\prime }}|\mathfrak{M}_{fi}|^{2}\propto \mathrm{tr}[\hat{V}%
\mathcal{\varrho }_{J}^{LS}(\mathbf{n},\mathbf{n}_{\ast })\hat{V}^{\dag }%
\mathcal{\varrho }_{J^{\prime }}^{L^{\prime }S^{\prime }}(\mathbf{n}^{\prime
},\mathbf{n}_{\ast }^{\prime })],  \label{probability}
\end{equation}%
where $\mathbf{n}_{\ast }$ and $\mathbf{n}_{\ast }^{\prime }$ are the arguments 
of the spherical tensors that appear in the complex conjugate matrix element.

In this letter, we present finite expansions of $\mathcal{\varrho }_{J}^{LS}(\mathbf{n},\mathbf{n}_{\ast })$ 
in powers of the scalar product
$\mathcal{S}^{\mu }\mathcal{L}_{\mu }$ and in terms of the polarization operators
$T_{LM}(\mathcal{S})$ and $T_{LM}(\mathcal{L})$. 
The latter are commonly employed to describe particle observables, such as magnetic dipole moments, 
electric quadrupole moments, etc. Our notation and normalization conventions 
follow those in Ref. \cite{Varshalovich:1988}.

A simple representation of $\mathcal{\varrho }_{J}^{LS}(\mathbf{n},\mathbf{n}%
_{\ast })$ can be constructed using a Frobenius covariant 
\begin{equation}
\mathcal{F}_{J}^{LS}=\prod\limits_{j\neq J}\frac{j(j+1) \mathcal{I}-\mathcal{J}^{2}}{%
j(j+1)-J(J+1)}\vspace{3mm}.  \label{Frobenius}
\end{equation}%
Here, $\mathcal{I}$ is the identity matrix of dimension $\left( 2S+1\right)
\times \left( 2S+1\right) $, and the product runs over
all values $j=|L-S|,\ldots ,L+S$, excluding $j=J$. 
In bra-ket notation, $\mathcal{F}_{J}^{LS}=\sum_{M}|JM\rangle\langle JM|$,
where the explicit dependence of the states $|JM\rangle$ on $L$ and $S$ is suppressed.

The addition theorem for
spherical harmonics allows to perform the summation of $\mathcal{\varrho }%
_{J}^{LS}(\mathbf{n},\mathbf{n}_{\ast })$: 
\begin{equation}
\sum_{J}\mathcal{\varrho }_{J}^{LS}(\mathbf{n},\mathbf{n}^{\prime })=\frac{%
2L+1}{4\pi }P_{L}(\mathbf{nn}^{\prime })\mathcal{I}.  \label{A2}
\end{equation}%
The Legendre polynomial multiplied by $(2L + 1)/(4\pi)$
projects onto the eigenspace associated with the orbital angular momentum eigenvalue $L$.
By applying $\mathcal{F}_{J}^{LS}$ to both sides of Eq. (\ref{A2}), we
obtain 
\begin{equation}
\mathcal{\varrho }_{J}^{LS}(\mathbf{n},\mathbf{n}^{\prime })=\mathcal{F}%
_{J}^{LS}\frac{2L+1}{4\pi }P_{L}(\mathbf{nn}^{\prime }),  \label{Major}
\end{equation}%
where the operator $\mathcal{L}_{\mu }$ entering $\mathcal{F}_{J}^{LS}$ acts on the vector $\mathbf{n}$. 

The finite expansion of $\mathcal{F}_{J}^{LS}$ in powers of $\mathcal{S}^{\mu
}\mathcal{L}_{\mu }$ displays a transparent and well‑defined structure.
The powers of $\mathcal{S}^{\mu }\mathcal{L}_{\mu }$, however, 
contain tensor operators formed as products of
$\mathcal{S}^{\mu_{1}}\ldots \mathcal{S}^{\mu _{K}}$ and $\mathcal{L}_{\mu _{1}}\ldots \mathcal{L}_{\mu _{K}}$, 
which are neither symmetric nor traceless, thereby obscuring the spin 
and orbital angular momentum content of the individual terms.

Spin polarization operators $T_{KM}(\mathcal{S})$ constitute a complete
basis in the space of $\left( 2S+1\right) \times $ $\left( 2S+1\right) $
Hermitian matrices for $0 \leq K \leq 2S$. 
They are irreducible tensor operators of the rotation group, offering a natural framework 
to describe interactions with external fields of specific multipolarity. 
In the subsequent discussion, the polarization operator $T_{KM}(\mathcal{L})$ 
associated with the orbital angular momentum will also be used; 
it is defined in the same way as $T_{KM}(\mathcal{S})$.

Under coordinate system rotations, the outer product $Y_{JM}^{LS}(\mathbf{n}%
)Y_{J^{\prime }M^{\prime }}^{L^{\prime }S\dag }(\mathbf{n}^{\prime })$
transforms via rotation matrices acting on the spin indices of the spin wave functions and the
coordinates of $\mathbf{n}$ and $\mathbf{n}^{\prime }$. The outer
product of the spin wave functions, $\chi _{S\sigma^{\prime}}\chi _{S\sigma }^{\dag }$, is
expandable in terms of the spin polarization operators \cite{Varshalovich:1988}. The decomposition
shows that the action of rotation matrices on the spin indices of $\chi _{S\sigma^{\prime}}$ and $\chi _{S\sigma }$
is equivalent to the action on the vector indices of the spin matrices, 
$\mathcal{S}^{\mu }$, and thus can be formally replaced by the latter. 

We present an expansion of $Y_{JM}^{LS}(\mathbf{n})Y_{J^{\prime }M^{\prime
}}^{L^{\prime }S\dag }(\mathbf{n}^{\prime })$ in terms of 
the polarization operators $T_{LM}(\mathcal{S})$ combined with the spherical harmonics $Y_{LM}(%
\mathbf{n})$ and $Y_{L^{\prime }M^{\prime }}(\mathbf{n}^{\prime })$, 
corresponding to the initial and final particle states.
The expansion enables to
present $\mathcal{\varrho }_{J}^{LS}(\mathbf{n},\mathbf{n}^{\prime})$ as a sum of the scalar products of 
$T_{LM}(\mathcal{S})$ and $T_{LM}(\mathcal{L})$, which act on a Legendre polynomial of degree $L$. 

Replacing the outer product
$\chi _{S\sigma^{\prime} }\chi _{S\sigma }^{\dagger}$ 
with a sum of the spin polarization operators
and summing over the magnetic quantum numbers
in the decomposition of $Y_{J_{1}M_{1}}^{L_{1}S}(\mathbf{n})Y_{J_{2}M_{2}}^{L_{2}S\dag }(\mathbf{n}^{\prime })$ 
yields
\begin{widetext}
\begin{align}
Y_{J_{1}M_{1}}^{L_{1}S}(\mathbf{n})Y_{J_{2}M_{2}}^{L_{2}S\dag }(\mathbf{n}%
^{\prime })& =\sum_{KL_{12}LM}(-1)^{-M_{2}+J_{2}-L_{2}+L-K-L_{12}}\Pi
_{KL_{12}J_{1}J_{2}}C_{J_{1}M_{1}J_{2}-M_{2}}^{LM}  \notag \\
& \times \left\{ 
\begin{array}{lll}
S & S & K \\ 
L_{2} & L_{1} & L_{12} \\ 
J_{2} & J_{1} & L%
\end{array}%
\right\} 
\{ T_{K}(\mathcal{S})\otimes \{ Y_{L_{1}}(\mathbf{n})\otimes Y_{L_{2}}(\mathbf{n}^{\prime })\} _{L_{12}}\} _{LM},
\label{YY}
\end{align}%
\end{widetext}
where 
$\Pi_{KL_{12}J_{1}J_{2}} = ((2K+1)(2L_{12}+1)(2J_{1}+1)(2J_{2}+1))^{1/2}$. The
summation is restricted by $0\leq K\leq 2S$ and is subject to 
the angular momentum triangle inequalities for $\left( L_{1},L_{2},L_{12}\right) $, $\left(
K,L_{12},L\right) $, and $\left( L,J_{1},J_{2}\right) $. 
The matrix element (\ref{Mfi}) with $\hat{V} = T_{KM}(\mathcal{S})$
can be obtained from
Eq.~(\ref{YY}) using orthonormality of the spin
polarization operators under the trace operation in the spin matrix space.

In the squared modulus of the matrix elements, the spherical tensors
are grouped into the pairs, as is evident from Eq.~(\ref{probability}). 
Accordingly, of particular interest is the special case $J_{1}=J_{2}=J$ and $L_{1}=L_{2}=L$, 
in which the summation is performed over $M_{1}$ for $M_{2}=M_{1}$. The result is represented in the form of Eq.
(\ref{Major}) with the Frobenius covariant given in terms of the
scalar product of $T_{KM}(\mathcal{S})$ and $T_{KM}(\mathcal{L})$:
\begin{equation}
\mathcal{F}_{J}^{LS}=(2J+1)\sum_{\substack{K=0}}^{2\min(S,L)}(-1)^{S+L+J+K}%
\sqrt{2K+1}\left\{
\begin{array}
[c]{lll}%
S & S & K\\
L & L & J
\end{array}
\right\}  \{T_{K}(\mathcal{S})\otimes T_{K}(\mathcal{L})\}_{00}. \label{Frobenius II}%
\end{equation}
The index $K$ takes values from zero up to the minimum of the two
maximum polarity values associated with the spin and orbital angular momentum polarization
operators. The operator $\mathcal{\varrho}_{J}^{LS}(\mathbf{n},\mathbf{n}^{\prime})$ 
is manifestly rotation-covariant.

The action of $\{T_{K}(\mathcal{S})\otimes T_{K}(\mathcal{L})\}_{00}$ on the function 
$(2L + 1)P_{L}(\mathbf{nn}^{\prime })/(4\pi)$ entering Eq.~(\ref{Major}) leads to the expression
\begin{align*}
(-1)^{K - L}\{T_{K}(\mathcal{S})\otimes\{Y_{L}(\mathbf{n})\otimes Y_{L}%
(\mathbf{n}^{\prime})\}_{K}\}_{00}. \label{rhoLSJTYY}%
\end{align*}

The simultaneous eigenspaces of $\mathcal{J}^{2}$ and $\mathcal{J}_{z}$ 
can be projected onto using an operator constructed as a composition of two Frobenius covariants:
\begin{equation}
\mathcal{F}_{JM}^{LS}= \mathcal{F}_{J}^{LS}
\prod\limits_{m\neq M}
\frac{m \mathcal{I} -\mathcal{J}_{z}}{m-M},
\end{equation}
where the product runs over all integers or half‑integers $m$ from $-J$ to $J$, excluding $m=M$. 
In bra-ket notation,
$\mathcal{F}_{JM}^{LS}=|JM\rangle\langle JM|$. The correspondence with
Villars' angular momentum projection operator $\mathcal{P}_{MM^{\prime}}^{J}=|JM\rangle \langle JM^{\prime}|$ commonly used in nuclear structure studies
\cite{Schmid:2004,Raduta:2015} is established by the identities
$\mathcal{P}_{MM}^{J}=\mathcal{F}_{JM}^{LS}$ and $\sum_{M}\mathcal{P}_{MM}^{J}=\mathcal{F}_{J}^{LS}$.
The off-diagonal matrix elements obey the relation
\begin{equation}
\mathcal{P}_{MM^{\prime}}^{J}=N_{JMM^{\prime}}\left(  \mathcal{J}_{\pm
1}\right)  ^{\left\vert M-M^{\prime}\right\vert }\mathcal{F}_{JM^{\prime}%
}^{LS},\label{rel}%
\end{equation}
where $\mathcal{J}_{\mu}$ is given in a cyclic basis. The upper sign $(+)$ is used when
$M>M^{\prime}$, and the lower sign $(-)$ is used otherwise. The normalization
constant $N_{JMM^{\prime}}$ is determined by the equation
\[
1/N_{JMM^{\prime}}={\prod\limits_{m}}\left(  \mp\sqrt{(J\pm m+1)(J\mp
m)/2}\right)  ,
\]
with the product taken over all values of $m$ within the interval
$[\min(M^{\prime},M \mp 1),\max(M^{\prime},M \mp 1)]$. 
The operator $\mathcal{P}_{MM^{\prime}}^{J}$ is computed numerically via 
integration over the rotation group space, whereas the algebraic expression given by 
Eq.~(\ref{rel}) relies on matrix multiplication.

Equation (\ref{Frobenius II}) can be reformulated in Cartesian tensor notation by expressing 
$\{T_{K}(\mathcal{S})\otimes T_{K}(\mathcal{L})\}_{00}$ as an operator proportional to
$\mathcal{O}_K = \mathcal{S}^{\mu_{1}}%
\ldots\mathcal{S}^{\mu_{K}}\mathcal{L}_{\nu_{1}}\ldots\mathcal{L}_{\nu_{K}}\theta_{\mu_{1}\ldots\mu_{K}}^{\nu_{1}\ldots \nu_{K}}$,
where $\theta_{\mu_{1}\ldots\mu_{K}}^{v_{1}\ldots v_{K}}$ 
is a symmetric, traceless tensor 
with numerous useful properties, including those of a projection operator
(see, e.g., \cite{Applequist:1989}). 
The symmetry under permutations of $\mathcal{L}_{\mu_{1}},\ldots ,\mathcal{L}_{\mu _{K}}$
enables the use of Fa{\`{a}} di Bruno's formula
to derive a double‑summation expression for $\mathcal{O}_K (2L + 1)P_{L}(\mathbf{nn}^{\prime })/(4\pi)$.

As such, $\theta_{\mu_{1}\ldots\mu_{K}}^{v_{1}\ldots v_{K}}$ arises as a multiplier 
in propagators of free particles with integer spins. 
A closely analogous role is assigned to $\mathcal{\varrho}_{J}^{LS}(\mathbf{n},\mathbf{n}^{\prime})$
in the description of spin‑$S$ particles in spherically symmetric potentials. 
The non-relativistic propagator for such particles can be decomposed as
\begin{align}
G_{S}(E,\mathbf{r},\mathbf{r}^{\prime})=\sum_{JL}
G_{JLS}(E,r,r^{\prime})\mathcal{\varrho}_{J}^{LS}(\mathbf{n},\mathbf{n}%
^{\prime}),
\end{align}
which explicitly separates the radial component  $G_{JLS}(E,r,r^{\prime})$ from the
angular part. Here, $E$ denotes the particle's energy, and the spatial
coordinates are defined by $\mathbf{r}=\mathbf{n}r$, $\mathbf{r}^{\prime
}=\mathbf{n}^{\prime}r^{\prime}$.
The explicit form of $\mathcal{\varrho}_{J}^{LS}(\mathbf{n},\mathbf{n}^{\prime})$
facilitates the treatment of 
higher-spin particles in spherically 
symmetric potentials.

The considered expansions of $\mathcal{F}^{LS}_J$ remain applicable in the relativistic setting after three-dimensional
reduction in a reference frame with spherical symmetry.

\vspace{2mm}

The author gratefully acknowledges Professor F. \v Simkovic 
for drawing attention to the importance of projection operator techniques
for electron capture theory.



\begin{thebibliography}{99}

\bibitem{Schmid:2004} K. W. Schmid, \textit{On the use of general symmetry-projected Hartree-Fock-Bogoliubov configurations in variational approaches to the nuclear many-body problem}, Prog. Part. Nucl. Phys. \textbf{52}, 565 (2004).
\bibitem{Raduta:2015} A. A. Raduta, \textit{Nuclear Structure with Coherent States},
(Springer International Publishing, Cham, Switzerland, 2015), pp. 523.

\bibitem{Berestetskii:1982} V. B. Berestetskii, E. M. Lifshitz, L. P.
Pitaevskii, \textit{Quantum Electrodynamics}. Vol. 4 (2nd ed.),
(Butterworth-Heinemann, 1982), pp. 680.

\bibitem{Cotogno:2020} S. Cotogno, C. Lorc\`{e}, P. Lowdon and M. Morales, 
\textit{Covariant multipole expansion of local currents for massive states
of any spin}, Phys. Rev. D \textbf{101}, 056016 (2020).

\bibitem{Krivoruchenko:1994} M. I. Krivoruchenko, \textit{Transitional
currents of spin 1/2 particles}, Usp. Fiz. Nauk \textbf{164}, 643 (1994)
[Phys. Usp. \textbf{37}, 601 (1994)].

\bibitem{Prokhorov:1999} L. V. Prokhorov, \textit{Infrared and collinear
divergences in gauge theories}, Usp. Fiz. Nauk \textbf{169}, 1199 (1999)
[Phys. Usp. \textbf{42}, 1099 (1999)].

\bibitem{Ahmadiniaz:2023} N. Ahmadiniaz, C. Lopez-Arcos, M. A. Lopez-Lopez,
Ch. Schubert, \textit{The QED four-photon amplitudes off-shell: Part 1},
Nucl. Phys. B \textbf{991}, 116216 (2023).

\bibitem{Simkovic:2021} F. \v Simkovic, \textit{Neutrino masses and interactions and neutrino experiments in the laboratory}, 
Phys. Usp. \textbf{64}, 1238 (2021). 

\bibitem{Krivoruchenko:2008} M. I. Krivoruchenko, Amand Faessler, \textit{%
Decays, contact P-wave interactions and hyperfine structure in $\Omega^{-}$
exotic atoms}, Nucl. Phys. A \textbf{803}, 173 (2008).

\bibitem{Varshalovich:1988} D. A. Varshalovich, A. N. Moskavlev, V. K.
Khersonsky, \textit{Quantum theory of angular momemtum}, (World Scientific
1988), pp. 528.

\bibitem{Applequist:1989} J. Applequist, \textit{Traceless cartesian tensor
forms for spherical harmonic functions: new theorems and applications to
electrostatics of dielectric media}, J. Phys. A: Math. Gen. \textbf{22},
4303 (1989).
\end{thebibliography}
\end{document}